# Probing Solar Polar Regions

Yuanyong Deng[1,2,27] Hui Tian[3,2] Jie Jiang[4] Shuhong Yang[1,2,27] Hao Li[2]
Robert Cameron[5] Laurent Gizon[5,6,7] Louise Harra[8,9]
Robert F. Wimmer-Schweingruber[10] Frédéric Auchère[11]
Xianyong Bai[1,2,27] Luis Bellot Rubio[12] Linjie Chen[2] Pengfei Chen[13,14]
Lakshmi Pradeep Chitta[5] Jackie Davies[15] Fabio Favata[16,17] Li Feng[18]
Xueshang Feng[2] Weiqun Gan[18,27] Don Hassler[19] Jiansen He[3]
Junfeng Hou[1,2,27] Zhenyong Hou[3] Chunlan Jin[1,2] Wenya Li[2]
Jiaben Lin[1,2,27] Dibyendu Nandy[20] Vaibhav Pant[21] Marco Romoli[22]
Taro Sakao[23,24] Sayamanthula Krishna Prasad[21] Fang Shen[2,28] Yang Su[18]
Shin Toriumi[23] Durgesh Tripathi[25] Linghua Wang[3] JingJing Wang[2]
Lidong Xia[26] Ming Xiong[2,28] Yihua Yan[2,1,27] Liping Yang[2]
Shangbin Yang[1,2,27] Mei Zhang[1,2,27] Guiping Zhou[1,2,27] Xiaoshuai Zhu[2]
Jingxiu Wang[1,2,27] Chi Wang[2]

1 (*National Astronomical Observatories, Chinese Academy of Sciences, Beijing 100101, People's Republic of China*)

2 (*State Key Laboratory of Solar Activity and Space Weather, National Space Science Center, Chinese Academy of Sciences, Beijing, 100190, People's Republic of China*)

3 (*School of Earth and Space Sciences, Peking University, Beijing 100871, People's Republic of China*)

4 (*School of Space and Earth Sciences, Beihang University, Beijing,102206, People's Republic of China*)

5 (*Max-Planck-Institut für Sonnensystemforschung, Justus-von-Liebig-Weg 3, 37077 Göttingen, Germany*)

6 (*Institut für Astrophysik, Georg-August-Universität Göttingen, 37077 Göttingen, Germany*)

7 (*Center for Astrophysics and Space Science, NYUAD Institute, New York University Abu Dhabi, Abu Dhabi, UAE*)

8 (*Physikalisch-Meteorologisches Observatorium Davos, World Radiation Center, 7260 Davos Dorf, Switzerland*)

9 (*Die Eidgenössische Technische Hochschule Zürich, 8093 Zürich, Switzerland*)

10 (*Institute of Experimental and Applied Physics, Kiel University, D-24118 Kiel, Germany)*

11 (*Institut d'Astrophysique Spatiale, 91405 Orsay, France*)

* 基金项目：国家重点研发计划项目（2022YFF0503800），国家自然科学基金项目（12425301, 12425305, 12473051）
投稿日期：
E-mail: dyy@nao.cas.cn, huitian@pku.edu.cn

12 (*Instituto de Astrofísica de Andalucía (IAA-CSIC), Apdo. de Correos 3004, 18080 Granada, Spain*)
13 (*School of Astronomy and Space Science, Nanjing University, Nanjing 210023, People's Republic of China*)
14 (*Key Laboratory of Modern Astronomy and Astrophysics, Ministry of Education, Nanjing 210023, People's Republic of China*)
15 (*RAL Space, STFC Rutherford Appleton Laboratory, Harwell Campus, Didcot OX11 0QX, UK*)
16 (*INAF—Osservatorio Astronomico di Palermo, Piazza del Parlamento, 1, 90134 Palermo, Italy*)
17 (*Department of Physics, Imperial College London, Exhibition Road, London SW7 2AZ, UK*)
18 (*Key Laboratory of Dark Matter and Space Astronomy, Purple Mountain Observatory, CAS, Nanjing 210023, China*)
19 (*Southwest Research Institute, Boulder, CO 80302, USA*)
20 (*Center of Excellence in Space Sciences India, Indian Institute of Science Education and Research Kolkata, Mohanpur 741246, India*)
21 (*Aryabhatta Research Institute of Observational Sciences, Nainital, 263002, India*)
22 (*Dip. di Fisica e Astronomia, Università di Firenze, Via Sansone 1, 50019 Sesto Fiorentino (FI), Italy*)
23 (*Institute of Space and Astronautical Science, Japan Aerospace Exploration Agency, Sagamihara, Kanagawa 252-5210, Japan*)
24 (*Department of Space and Astronautical Science, School of Physical Sciences, SOKENDAI (The Graduate University for Advanced Studies), Sagamihara, Kanagawa 252-5210, Japan*)
25 (*Inter-University Centre for Astronomy and Astrophysics, Post Bag 4, Ganeshkhind, Pune 411007, India*)
26 (*School of Space Science and Technology, Shandong University, Weihai, Shandong 264209, People's Republic of China*)
27 (*School of Astronomy and Space Science, University of Chinese Academy of Sciences, Beijing 100049, People's Republic of China*)
28 (*College of Earth and Planetary Sciences, University of Chinese Academy of Sciences, Beijing 100049, People's Republic of China*)

**Abstract** The magnetic fields and dynamical processes in the solar polar regions play a crucial role in the solar magnetic cycle and in supplying mass and energy to the fast solar wind, ultimately being vital in controlling solar activities and driving space weather. Despite numerous efforts to explore these regions, to date no imaging observations of the Sun's poles have been achieved from vantage points out of the ecliptic plane, leaving their behavior and evolution poorly understood. This observation gap has left three top-level scientific questions unanswered, 1) How does the solar dynamo work and drive the solar magnetic cycle? 2) What drives the fast solar wind? 3) How do space weather processes globally originate from the Sun and propagate throughout the solar system? The Solar Polar-orbit Observatory (SPO) mission, a solar polar exploration spacecraft, is proposed to address these three unanswered scientific questions by imaging the Sun's poles from high heliolatitudes. In order to achieve its scientific goals, SPO will carry six remote-sensing and four in-situ instruments to measure the vector magnetic fields and Doppler velocity fields in the photosphere, to observed the Sun in the extreme ultraviolet, X-ray, and radio wavelengths, to image the corona and the heliosphere up to 45 $R_\odot$, and to perform in-situ detection of magnetic fields, and low- and high-energy particles in the solar wind. The SPO mission is capable of 1) providing critical vector magnetic fields and Doppler velocities of the polar regions to advance our understanding of the origin of the solar magnetic cycle, 2) providing unprecedented imaging observations of the solar poles alongside in-situ measurements of charged particles and magnetic fields from high heliolatitudes to unveil the mass and energy supply that drive the fast solar wind, and 3) providing observational constraints for improving our ability to model and predict the three-dimensional (3D) structures and propagation of space weather events.

# 0 Introduction

## 0.1 Forum Overview

The poles remain as the least-understood and most mysterious territory on the Sun, yet they are essential for comprehending the Sun's magnetic dynamics and their impact on solar activities and space weather phenomena. The proposed Solar Polar-orbit Observatory (SPO)[1], with an orbital inclination angle of 80° and a small ellipticity, is designed to make breakthroughs on three top-level scientific questions, i.e. 1) How does the solar dynamo work and drive the solar magnetic cycle? 2) What drives the fast solar wind? 3) How do space weather processes globally originate and propagate throughout the solar system? In order to refine the scientific objectives and maximize the scientific output of the SPO mission, the International Space Science Institute - Bejing (ISSI-BJ) organized an international forum in Beijing during 7-8 November 2024. During this forum, the preliminary scientific objectives of the SPO mission and the tentatively selected payloads were presented. This forum also served as a platform for discussions on how the SPO observations could advance our understanding of these most important scientific questions in solar physics. Additionally, suggestions and international collaborations regarding the instrument design, science preparation and synergies with other missions have also been explored. SPO is scheduled for launch in January 2029. It will, for the first time, image the Sun's poles from high heliolatitudes. The achieved observations are expected to provide invaluable information to advance our knowledge in the solar physics.

## 0.2 Importance of Exploring the Sun's Poles

The Sun is the driving force behind life on Earth, providing the energy, warmth, and light necessary to sustain our planet's habitability. No aspect of life was unaffected by the Sun[2]. Scientific observations of the Sun began over 400 years ago, and have explored nearly the entire electromagnetic spectrum, including X-rays, ultraviolet, optical, infrared, and radio radiation, as well as the solar neutrinos. Despite these advance, to date, we still lack a detailed view of the Sun's poles and their behavior, simply because nearly all the observations of the Sun are limited in the vicinity of the ecliptic plane, which significantly hinder our understanding of the solar dynamo, and the origin of the fast solar wind.

Exploring the Sun's poles is of great importance in solar physics, since the magnetic fields and the dynamical processes in these regions play a key role in characterizing the evolution of the solar internal dynamo and the distribution of the external heliospheric magnetic fields[3]. In addition, the polar magnetic fields likely serve as key indicators of the strength of the next solar cycle[4,5], and dominate the large-scale coronal structures[6]. Moreover, the polar coronal holes are the primary source regions for the persistent and stable fast solar wind[7]. The absence of imaging observations from high latitudes leaves the key questions about the solar dynamo and fast solar wind unresolved. In addition, observations from a polar vantage point, combined with those obtained in the ecliptic plane, will be able to provide comprehensive boundary conditions for data-driven 3D heliospheric magnetohydrodynamic modeling, laying the foundation for understanding the global structures of the heliosphere and predicting the space weather phenomena in the heliospheric space.

# 1 Observations of the Solar Polar Regions

Although until to date the Sun's poles have not yet been directly imaged, several efforts have been made to explore these regions. In this section, we briefly summarize the existing observations of the polar regions obtained both in and out of the ecliptic plane.

## 1.1 Observations in the Ecliptic Plane

Most observations of the Sun's poles are obtained from vantage points near the ecliptic plane. Since the tilt angle of the Sun's rotation axis is approximately 7.25° with respect to the normal of the ecliptic plane, during specific time periods, the polar region is observable even in the ecliptic plane although with an oblique angle larger than 82°. For instance, the south and north poles become observable from the Earth's direction in March and September, respectively.

Observations of the magnetic fields in the polar regions date back to the 1950s, when Babcock et al.[8] reported a mean field strength of 1 G. Deng et al.[9] later used a videomagnetograph installed at Huairou Solar Observing Stations[10] to infer the vector magnetic fields from the polarization observations of the Fe I line at 532.4 nm. Their results showed that during the minimal phase between the solar cycle 22 and 23, the net flux in the south polar region was approximately $-2.5 \times 10^{22}$ Mx, which is around the order of the interplanetary magnetic field measured at a distance of 1 AU[9].

The Full Stokes observations of the polar regions with higher spatial and spectral resolutions were achieved since the launch of the Hinode satellite[11], which carries the Stokes Spectro-Polarimeter (SP) attached to the Solar Optical Telescope (SOT)[12].

Applying a Milne-Eddington inversion code, the landscape of the vector magnetic fields in the polar region was obtained by Tsuneta et al.[13], revealing vertically oriented magnetic flux tubes with a strength of 1 kG in the polar region between latitudes of 70° and 90°. In addition, ubiquitous horizontal magnetic fields were also identified in these regions. During the solar minimum the ratio between positive and negative magnetic fluxes is 1/2 in polar regions, which suggests only 1/3 of these fluxes remain open[14]. Further studies by Jin et al.[15] identified more than 300 small-scale bipolar magnetic emergences (BMEs) in the polar region. The magnetic axes for these BME are randomly distributed, which does not follow the Hale's or the Joy's laws typically for the active regions

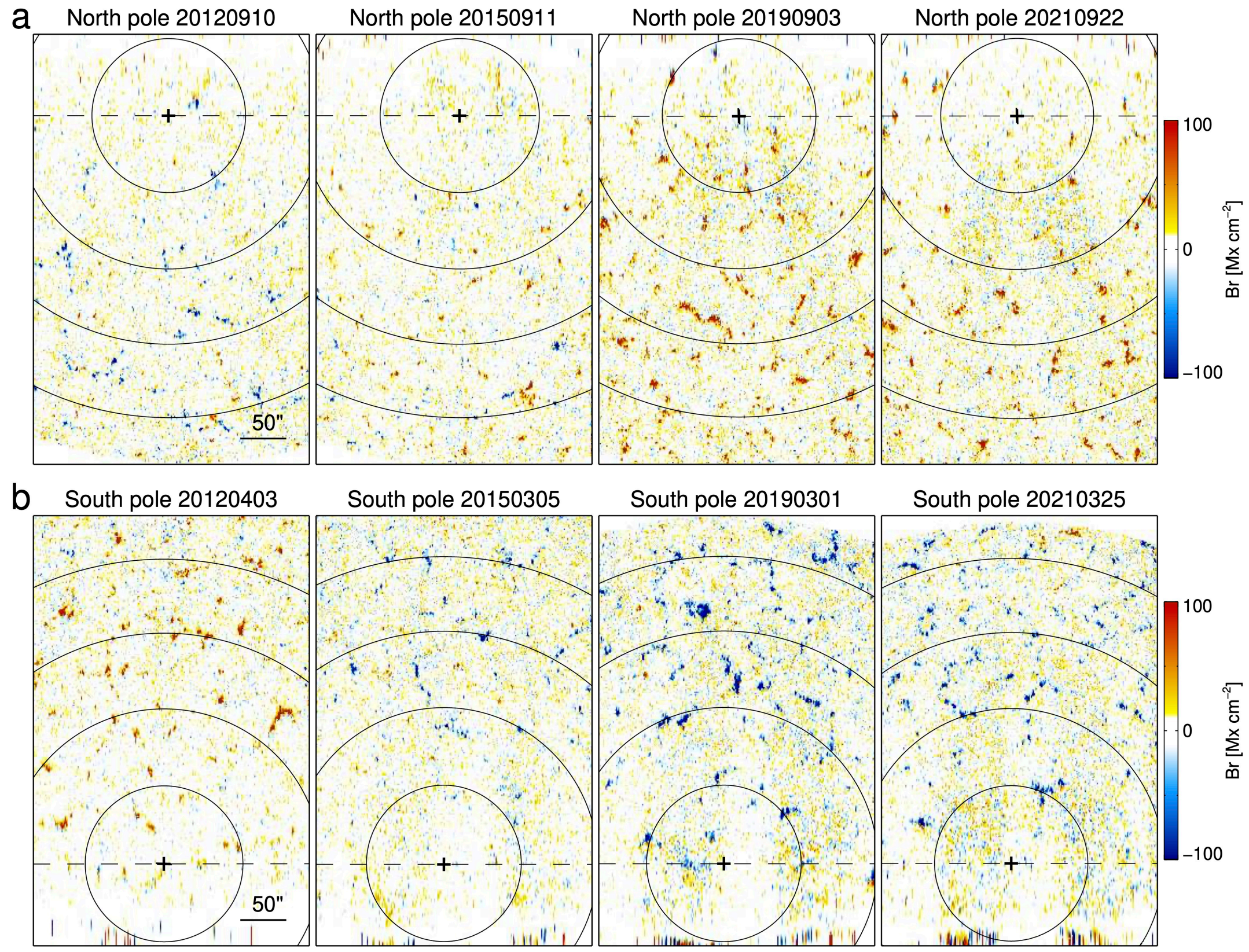

Figure 1 Polar view of the radial magnetic flux distribution in the polar caps observed by Hinode. (a) Radial magnetic field in the north polar cap measured in 2012, 2015, 2019, and 2021. (b) Similar to (a) but for the south pole. The plus signs mark the poles, and the solid curves indicate the latitude separated by 5◦. Figure from Yang et al[16].

Using the polar observations obtained by Hinode/SOT-SP from 2012 to 2021, covering the pre-reversal phase of the 24th solar cycle and the early rising phase of the 25th solar cycle, Yang et al.[16] studied the long-term variation of the polar magnetic fields (see Figure 1), and identified that the polarity reversal at high latitude lagged behind that at low latitude, indicating a poleward magnetic flux migration. Additionally, Yang et al.[17] found that the radial magnetic flux densities are weaker at high latitudes than those at lower latitudes during the solar cycle except during the polarity reversal phase. To explore the meridional flow behind these observational evidences, a surface flux transport model[18] was employed by Yang et al.[17] to simulate the global radial magnetic field, revealing a poleward meridional flow located between the latitudes of 0° and 70° with a maximum amplitude of 11 m $s^{-1}$ at 35° latitude, and an equatorward meridional flow located between the latitudes of 70° and 90° with a maximum amplitude of 3 m $s^{-1}$ at 80° latitude.

In addition to the photospheric magnetic fields, observations of the upper atmosphere in the polar regions are also crucial. Coronal holes, regions where the open magnetic field lines extend into the interplanetary space, appear dark in extreme ultraviolet (EUV) and X-ray wavelengths and serve as primary source regions of the fast solar wind. Observations of coronal holes over the last two decades have revealed numerous small-scale apparent collimated flows through the solar atmosphere. These include solar chromospheric spicules, transition region network jets and EUV propagating intensity disturbances termed jetlets, in plume regions[19−21] These features are likely driven by magnetic inter change reconnection operating on small spatial scales of a few 100 km in the solar atmosphere[22−25], triggered by rapidly changing photospheric magnetic field on timescales less than 5 minutes[26].

The X-ray jets observed by Hinode have revealed two distinct velocities of approximately 200 and 800 km $s^{-1}$. The large number of X-ray jets, coupled with the high velocities of the apparent outflows, suggests they may contribute to the fast solar wind[27]. Furthermore, in addition to jets observed in plume regions, Tian et al.[28] reported repetitive jets at a temperature of million degrees in the inter-plume regions. The role of these small jet features and their direct contribution to the solar wind remains a topic of ongoing debate. In addition to the jets, spicules permeated in the chromosphere are also considered as potential mass contributors to the solar wind. Alfvénic waves carried by spicules with sufficient energy flux to accelerate the solar wind was reported by De Pontieu et al.[29] and McIntosh et al.[30]. However, their precise contribution remains an open question in heliophysics.

The Solar Ultraviolet Measurements of Emitted Radiation (SUMER) instrument aboard the Solar and Heliospheric Observatory (SOHO) mission[31] provided high resolution spectroscopic observations at different locations on the Sun. Hassler et al.[32] explored a mid-latitude quiet Sun region and a polar region using the Ne VIII 77nm emission line. They showed that in the mid-latitude region, the strongest upflowing plasma occurred at the edges of the convective cells, while at the poles, the plasma was predominantly blue-shifted[33,34]. Spectroscopic measurements of the poles are also performed regularly by the Hinode mission. The observations suggest that the nascent fast solar wind in the coronal holes originates in the transition region and propagates along the open field lines[35]. A study of the polar regions over 6 years was carried out by Harra et al.[36], revealing enhanced line width in the coronal Fe XII emission lines known as non-thermal velocity. Additionally, SUMER observations of various ions suggest that these ions are heated by ion-cyclotron waves, which are generated through mode conversion from Alfvén waves[37] .

## 1.2 Observations out of the ecliptic plane

The above-mentioned results are all obtained via limb observations due to the observational limitations in the ecliptic plane. The observations of the polar regions out of the ecliptic plane remain rare, but notable exceptions include the Ulysses spacecraft[38] and the ongoing Solar Orbiter mission[39], which is gradually leaving the ecliptic plane.

The Ulysses spacecraft, by using a Jupiter fly-by, was the first mission to achieve a large inclination orbit of around 80° with respect to the ecliptic plane. This unique trajectory allowed Ulysses to pass over the solar poles multiple times during its mission, completing three polar orbits (each covering 6 years) over different phases of Solar Cycles 22 and 23[40]. Although Ulysses did not carry imaging instruments, its in-situ instruments provided crucial measurements of charged particles and magnetic fields as close as 1.34 AU from the Sun. These data offered unprecedented insights into the structure and dynamics of the heliosphere from high heliolatitudes.

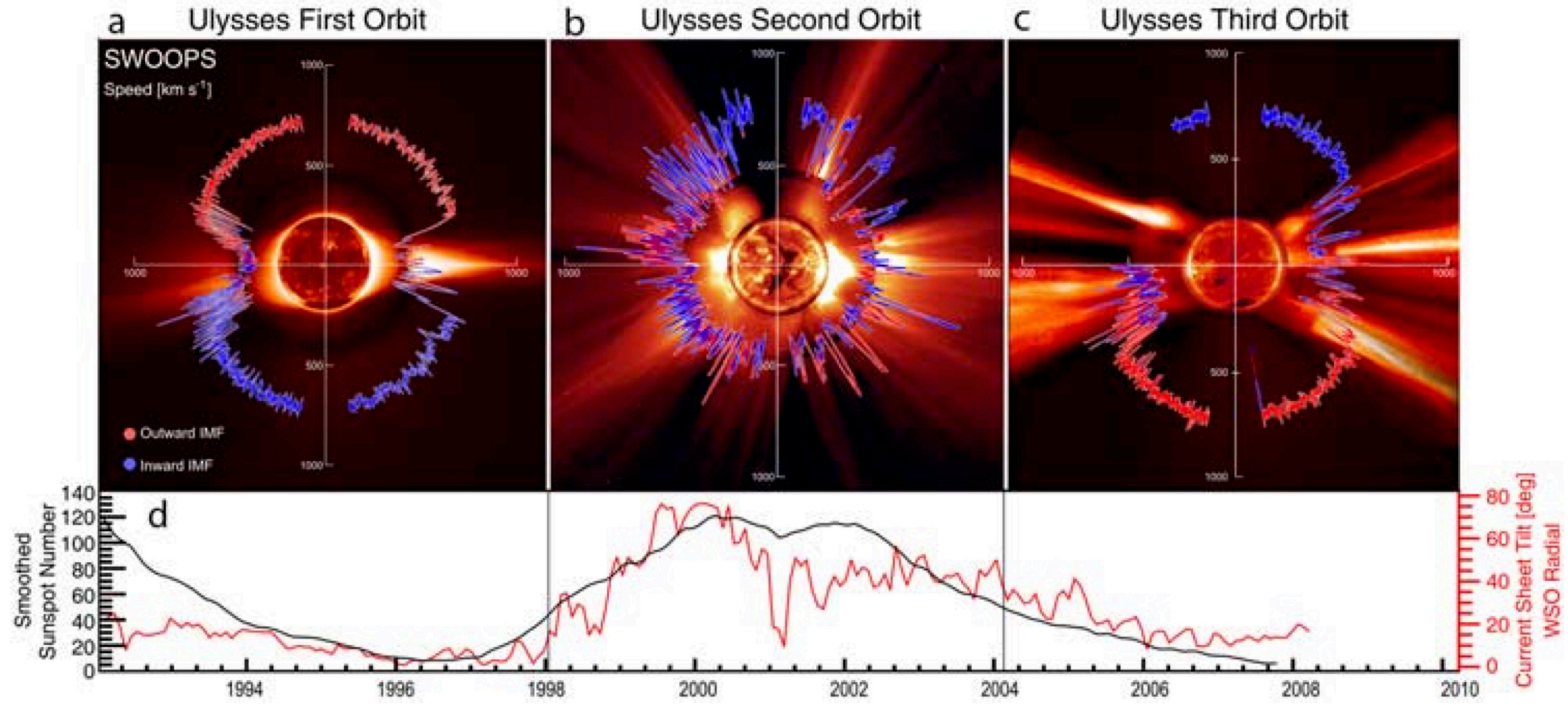

Figure 2 (a)–(c) Polar plots of the solar wind speed, colored by the Interplanetary Magnetic Field (IMF) polarity for Ulysses' three polar orbits. (d) Contemporaneous values for the smoothed sunspot number (black) and heliospheric current sheet tilt (red). Figure from McComas et al[41].

As seen in Figure 2, the Ulysses' first orbit covered a solar minimum, revealing that the solar wind speeds are higher at the poles than at the equator. Its second orbit detected a more complex global structure of the solar wind across all heliolatitudes[41]. The fast solar wind emerging from polar coronal holes was measured at speeds of approximately 750 km $s^{-1}$. As the Sun rotates, the fast wind streams from equator-ward extensions of the polar coronal holes interacts with the slower wind streams from the equatorial streamer belt, forming corotating interactions regions (CIRs) due to the Sun's rotation. Ulysses' orbit allowed for extensive CIR observations, which greatly enhanced our understanding the three-dimensional structure of the heliosphere. These findings were summarized in an ISSI publication on CIRs[42].

The solar wind flows away from the Sun, carrying with it both the frozen-in magnetic field and a tracer of the coronal region from which it originates. The elemental and charge-state composition of the heavy ions in the solar wind is determined in the chromosphere and corona, and remains unchanged thereafter[43]. This composition serves as an excellent discriminator for identifying the solar wind origin, even in highly compressed and heated plasma within CIRs[44].

One of the most remarkable discoveries made by Ulysses was that energetic particles accelerated at CIR-driven shocks are modulated by solar wind streams up to very high latitudes. Two mechanisms were proposed to explain this unexpected behavior. In the more traditional one, random motions of the photospheric footpoints of magnetic field lines "braid" the interplanetary field. Reconnection and enhanced turbulence of the field over the solar magnetic poles also play a role in this mechanism. Together with a small amount of scattering of particles across field lines, this mechanism is capable of explaining the unexpected Ulysses observations. The other, less traditional, mechanism involved a systematic latitudinal transport of magnetic field lines. Together, the photospheric differential rotation, the superradial expansion of coronal hole wind and the rigid rotation of coronal holes can result in significant latitudinal spread of the heliospheric magnetic field. Fisk et al.[45] provided an in-depth discussion of the commonalities and differences of these two mechanisms.

The Solar Orbiter mission from the European Space Agency, launched in 2020, is designed to make close approaches to the Sun, within Mercury's orbit, every 150 days. The mission will use a series of Venus Gravity Assists (VGAs) to gradually raise the spacecraft orbit out of the ecliptic plane to a maximum heliolatitude of 34°. The Solar Orbiter carries six remote-sensing instruments including the Extreme Ultraviolet Imager (EUI)[46], the Spectral Imaging of the Coronal Environment (SPICE)[47], the Polarimetric and Helioseismic Imager (PHI)[48], the Spectrometer/Telescope for Imaging X-rays (STIX)[49], a coronagraph Metis[50] and the Solar Orbiter Heliospheric Imager (SoloHI)[51], and four in-situ instruments including the Solar Wind Analyser (SWA)[52], the Radio and Plasma Waves (RPW) instrument[53], the Magnetometer (MAG)[54] and the Energetic Particle Detector (EPD)[55].

Although the observations obtained so far are still limited to the ecliptic plane, Solar Orbiter has already demonstrated its capability to provide invaluable insights into the Sun. The EUI has pushed the boundaries of the spatial resolution in the EUV imaging and has discovered small-scale transient brightenings, known as "campfires"[56]. Its High-Resolution Imager (HRI) allows scientists to observe fine structures in the solar atmosphere, capturing details as small as 200 km when the spacecraft is at its closest approach of 0.29 AU[57]. By combining the far-side magnetograms obtained by PHI with the magnetograms obtained by SDO/HMI, Loeschl et al.[58] produced multi-view synoptic magnetic field maps. Furthermore, the coordinated in-situ measurements by Solar Orbiter and the Parker Solar Probe (PSP) shows that the damping and mechanical work performed by the Alfvén waves are sufficient to power the heating and acceleration of the fast solar wind in the inner heliosphere[59]. As Solar Orbiter gradually increases its heliolatitude to 34°, it is expected to provide increasingly valuable data to complement high heliolatitude solar observations.

# 2 Top-Level Scientific Questions Regarding the Solar Polar Regions

Despite ongoing efforts to explore the Sun's polar regions, direct imaging observations of the Sun's poles from high heliolatitudes remain unavailable, leading to three unsolved Sun's mysteries primarily due to our limited understanding of the physical conditions in the solar polar regions[60].

## 2.1 How Does the Solar Dynamo Work and Drive the Solar Magnetic Cycle?

### 2.1.1 Key unresolved issues regarding the solar magnetic cycle

The solar magnetic cycle refers to the periodic variation in sunspot numbers on the solar surface, typically spanning approximately 11 years. This cycle governs the Sun's overall magnetic activity, encompassing phenomena such as sunspots, solar flares, and coronal mass ejections (CMEs). During each cycle, the Sun's magnetic poles undergo a reversal, with the magnetic polarities of the north and south poles switching.

The Sun's global magnetic fields are generated through a dynamo process. The solar differential rotation acts upon the poloidal field to generate the toroidal field, which manifests at the solar surface as active regions. The meridional circulation then transports the magnetic fluxes of the following polarity fields in the active regions towards the polar regions, regenerating the poloidal field and leading to a polarity reversal of the polar magnetic field[61]. The process repeats itself, resulting in the solar cycle (see Figure 3)[62].

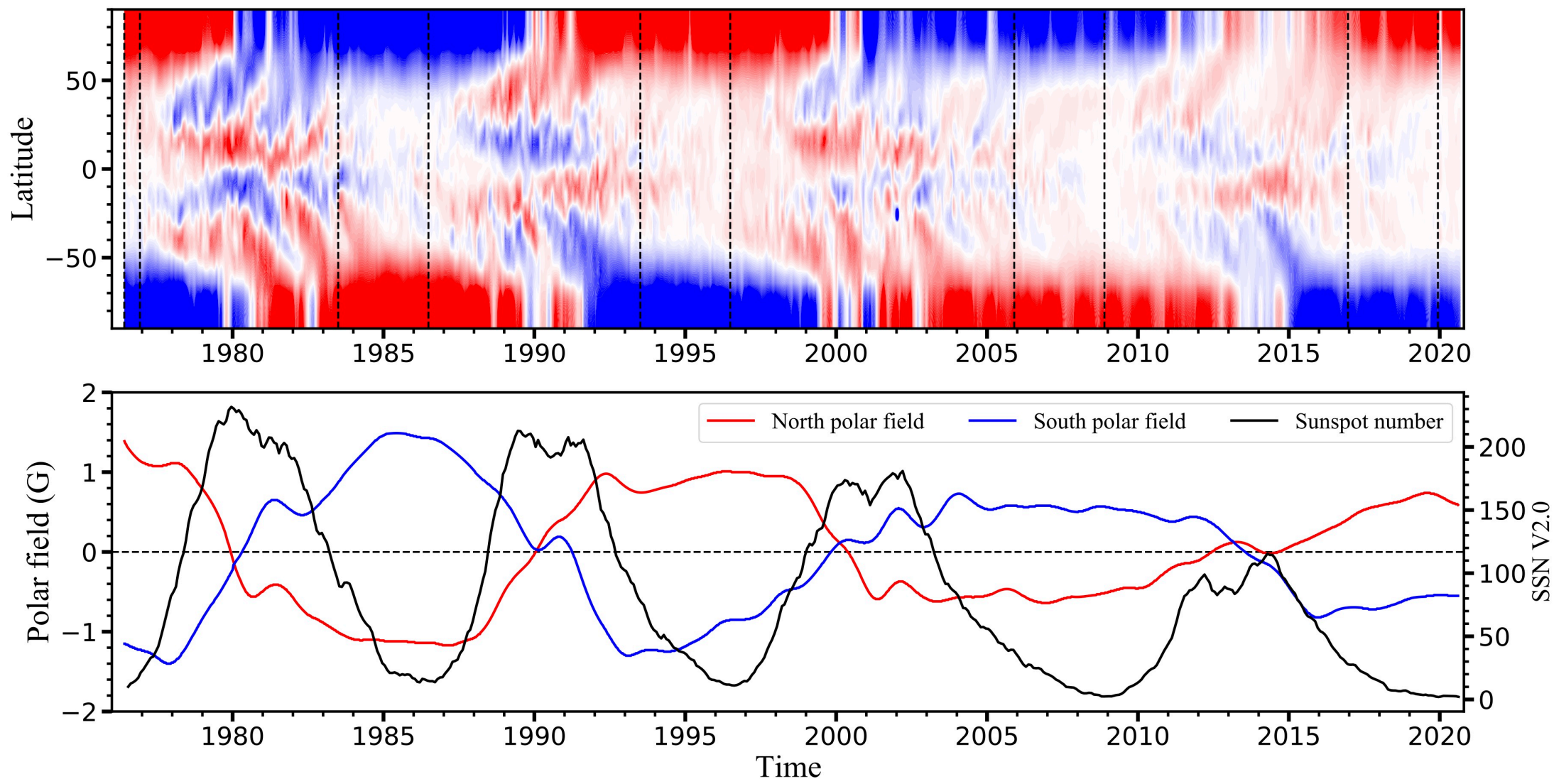

Figure 3 Surface magnetic field evolution of the Sun since 1976. (a) The butterfly diagram of magnetic evolution. (b) The time evolution of sunspot number (black line), and that of the magnetic intensity in the south (red line) and the north (blue line) poles. Figure from Guo et al[61].

During the past decades, the flux transport dynamo (FTD) developed by Wang et al.[63] and Choudhuri et al.[64] works as a paradigm for understanding the solar cycle. A key prerequisite of the dynamo model is that the meridional flow at the base of the convection zone must be equatorward. If there is either no flow or a poleward flow at the bottom of the convection zone, this dynamo model fails to reproduce the observed solar behavior[65]. However, some helioseismology inversions yielded conflicting results about the meridional flow, especially in deeper regions below about $0.9R_{\odot}$. Some studies suggest negligible flow[66] or even a poleward flow[67] at the bottom of the convection zone, preventing the classical FTD models that rely on the tachocline from functioning properly. Hence, this raises key questions to be addressed by future measurements: What is the realistic profile of the meridional flow, and if the flow is indeed poleward, how does the solar dynamo operate? Especially, where is the toroidal field originated and what mechanism generates the poloidal field.

The location, where the toroidal magnetic fields generate, is one of the most debated issues in the solar dynamo theory. Traditionally, it was believed that the toroidal field forms in the tachocline, a shear layer at the base of the convection zone[68], leading to the development of traditional dynamo models. However, several challenges have emerged against this view: 1) The magnetic field characteristics in fully convective stars are similar with those in stars that possess a tachocline. 2) Traditional dynamo models tend to generate strong toroidal fields at high latitudes. 3) These models often produce quadrupolar magnetic field features that conflict with observations. 4) The meridional flow at the bottom of the convection zone is assumed to be equatorward. However the real form of meridional flows inside the Sun remains highly debated. These challenges have prompted the development of a new generation of dynamo models, proposing that the toroidal field is generated within the convection zone rather than the tachocline, in an effort to unveil the origin of the solar magnetic cycle. In the dynamo model developed by Zhang et al.[69], the poloidal field is dominated by large-scale dipolar and octupolar components, consistent with the magnetic power spectrum derived from observations by Luo et al.[70]. This configuration suggests that the radial shear plays a negligible role in the generation of the toroidal field, and that the tachocline has a minimal role in its generation. Instead, the toroidal field is likely generated within the bulk of the convection zone. While these new models address many of the shortcomings of traditional theories, further validation is needed through observations and simulations.

Differential rotation and its temporal variations, referred to as torsional oscillations (TO), are also critical for understanding the solar dynamo. So far detailed measurements of TO near the solar poles are still lacking[71]. Torsional oscillations display both poleward and equatorward branches originating from about ±50° latitudes. Although the poleward branches are more pronounced, they lack of detailed observations. The properties of TOs, especially the poleward branches could provide constraints on the configurations of magnetic field in the convection zone and further help to pinpoint the location of the toroidal field generation[72]. Currently, there are controversial results about torsional oscillations. Kosovichev et al.[73] found that at low latitudes, the acceleration bands first appear at the bottom of the convection zone and rise toward the near surface, suggesting that the seat of the solar dynamo is primarily located in a high-latitude zone of the tachocline. In contrast, the helioseismology results given by Vorontsov et al.[74] showed that the acceleration bands first appear near surface and propagate downward with time, leading Vasil et al.[75] to propose a surface origin of the solar dynamo.

The mechanism for poloidal field generation also remains a topic of ongoing debate. Parker et al.[76] first proposed that the poloidal field arises from the kinematic helicity of turbulent convection, supported by mean-field theory. However, this theory has not yet been confirmed by observations. In contrast, the Babcock-Leighton (BL) mechanism[77–79], proposed in the 1960s based on observations, has gained increasing observational support in recent years and is now widely accepted. Cameron et al. [3] have demonstrated that surface polar field is the major source of the toroidal flux, which means that the BL mechanism is the principal contributor to the poloidal field. So far the proposed ingredients affecting the polar field evolution include 1) active region emergence as the flux source; 2) surface flux transport parameters including meridional flow and turbulent diffusion; 3) subduction of the polar field by the meridional flow sinking underneath the surface; and 4) radial diffusion occurring across the surface. The roles of the latter two ingredients in the generation of the polar field require further investigation.

In addition, the solar polar magnetic field participates in the solar global dynamo process and may serve as a seed field for the subsequent solar cycle, characterizing the solar dipole magnetic field, which corresponds to the poloidal magnetic field[80]. Studies have shown that the polar magnetic field during the solar cycle minimum phase has a strong correlation with the strength of the subsequent solar cycle[4,5]. Bhowmik developed a novel method for predicting the strength of the solar polar field at cycle minimum, which was then used as input to forecast the strength and timing of the next solar magnetic cycle[81]. Their prediction indicated that solar magnetic cycle 25 would be similar or slightly stronger in strength relative to the cycle 24. The confirmation of this prediction supports that the polar magnetic field could act as a precursor for predicting strength of the solar magnetic cycle[82,83]. Imaging the polar magnetic fields from high latitude will provide more accurate constraints for data-driven predictions of the solar magnetic cycle[84].

Turbulent convection is the small-scale flow field, which is essential for the dynamo behavior. Due to the solar rotation, the poles provide a unique window for revealing the dynamical regime and parameters of the convection zone. Through a series of numerical simulations, Hindman et al.[85] explored how the nature of convective structures, especially the polar view, transitions through a series of morphological regimes as the Rossby number varies. Gizon et al.[86] reported observations of a rich spectrum of inertial modes of the Sun over a wide range of latitudes, particularly the high-latitude m = 1 mode. Figure 4 shows the amplitude of the high-latitude m = 1 mode measured over the last two solar cycles[87] These inertial modes have diagnostic potential for the latitudinal entropy gradient, super-adiabaticity, and turbulent viscosity. These modes could also perhaps transport magnetic flux across the poles, and need to be taken into account to properly evaluate how much net toroidal magnetic flux is produced each cycle. The difficulty is that the flows close to the poles are poorly constrained from the ecliptic. Observations from a polar viewpoint can fill this gap. Moreover, the high-latitude inertial modes can also be used as diagnostics to improve our understanding of the structure of the solar convection zone[86].

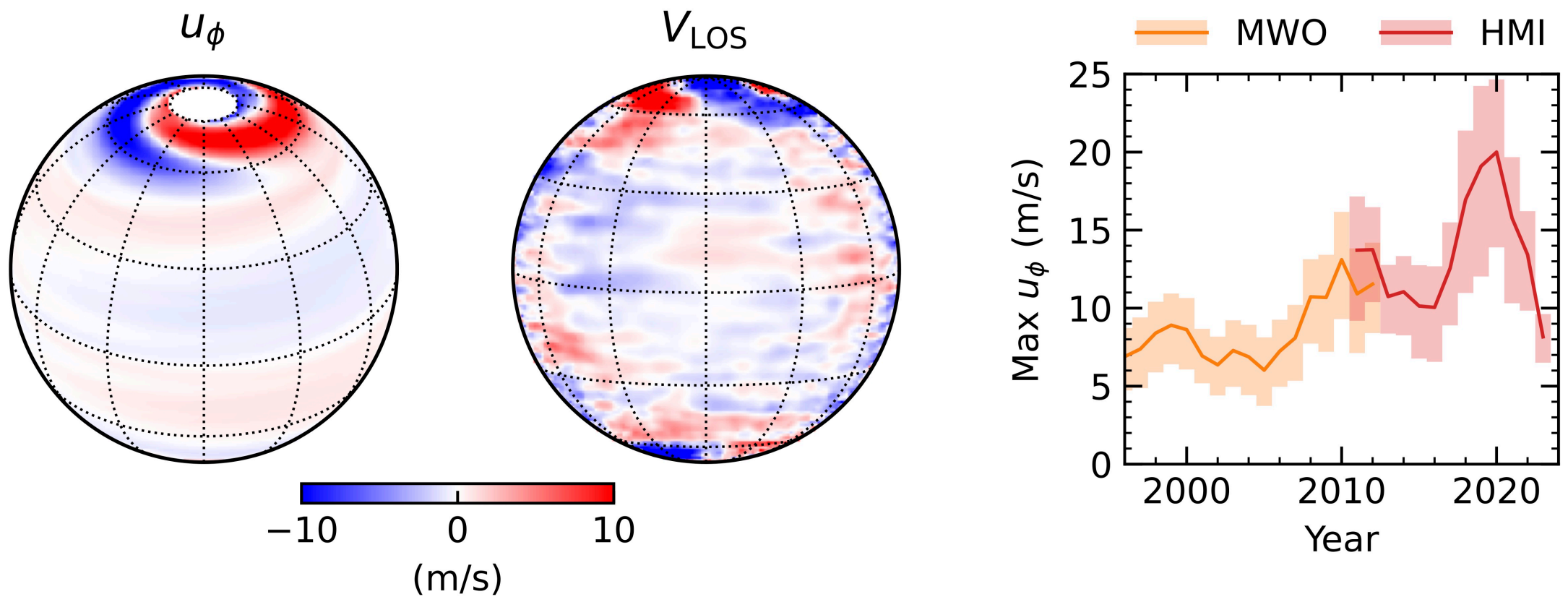

Figure 4 Observations of the high-latitude m = 1 solar inertial mode, which propagates retrograde in the Carrington frame (frequency −86 nHz). The left panel displays the mode's longitudinal velocity, $u_{\varphi}$, derived from ring-diagram helioseismology applied to HMI observations[86]. The middle panel presents the mode's line-of-sight velocity, as seen in filtered HMI Dopplergrams. The right panel is a plot of the mode's maximum velocity over time, using data from both HMI and Mount Wilson Observatory. The amplitude of the mode is larger around solar cycle minima, as noted by Liang & Gizon[87].

### 2.1.2 Limitations in current research

Certain limitations currently hinder progress in addressing the key unresolved issues of the solar cycle outlined in the previous subsection. One major challenge is the limited understanding of the precise characteristics of meridional circulation within the Sun. The radial structure of the meridional flow is critical for constraining our understanding of the solar convection, however, the flows are weak and the waves propagate quickly at this depth. Observations of the meridional circulation, especially in the deeper layers, show different results depending on the analytical method used, and a consensus has yet to be reached**.** This discrepancy is partly due to the limited observations from a single vantage point in the ecliptic plane.

Helioseismology enables the detection of solar internal flows by analyzing the propagation of pressure waves through the Sun. Local helioseismic techniques, which track and analyze the Doppler velocity maps of small regions on the solar surface, can infer internal flows beneath the Sun's surface. Nevertheless, due to the limited viewing direction from the ecliptic plane, current observations are insufficient for studying the polar regions using local helioseismology. The projection effect, center-to-limb effect and shrinking-Sun effect arising from the oblique observations introduce significant errors, that hinder the velocity measurements in local helioseismology[88], making the inversion of the high-latitude flow fields even more uncertain.

The distribution of polar magnetic flux presented in the magnetic butterfly diagram largely depends on the theoretical extrapolation of the solar surface magnetic flux transport model, since the measurement of the polar magnetic field strength is still a challenging task due to the observational projection effect. When observing from the ecliptic plane, the predominantly radial polar magnetic field appears as a transverse component, complicating its measurement. Additionally, the oblique viewing angle causes a single pixel to correspond to a larger spatial scale in the polar regions, reducing the spatial resolution and giving rise to the difficulties in the small-scale magnetic field measurements. Observations near the solar limb also are also affected by the limb darkening effect, which alters spectral line intensities and formation heights, thereby complicating magnetic field measurements.

### 2.1.3 The need for polar observations

To overcome the limitations outlined above, a spacecraft with a high inclination orbit with respect to the ecliptic plane is needed to perform direct imaging of the solar polar regions from high latitudes. Such observations would provide essential observations of the magnetic fields and internal flows in these critical regions, which are critical for unveiling the origin of the solar magnetic cycle and for pinpointing the solar dynamo model.

Observations of the Doppler velocities in the polar regions from a high heliolatitude vantage point, combined with the observations in ecliptic plane are the effective tools to get the differential rotation and convection structures near the poles and meridional flow in the deep convection zone[89]. Multi-vantage-point observations significantly improve spatial coverage, enhancing the signal-to-noise ratio[90] by mitigating the shrinking-Sun and center-to-limb effects. They also reduce the spatial leakage in the spectrum, and improve the spatial resolution of the sensitivity kernel in helioseismic inversion[91].

Vector magnetic field measurements in the polar region are crucial to determine whether the polar field is dominated by the radial component. With tilt quenching[92,93] and latitudinal quenching[94], no radial diffusion is required to get the regular polar field reversal in the surface flux transport simulations for cycles 15-21[95]. Similarly, in a 2D BL dynamo model, Cameron et al.[96] demonstrated that radial boundary condition combined with near surface radial pumping can obtain the surface field evolution consistent with results from SFT simulations without radial diffusion. Luo et al. (2024, in prep.) present a more general formulation of no radial diffusion boundary condition. However, whether radial diffusion actually occurs remains an open question. The measurement of the net polar flux and verification of the net flux across the equator could help to answer the question.

In addition, our understanding of the solar dynamo would also be enhanced by synoptic measurements of the latitudinal component of the solar large-scale magnetic field. Currently, synoptic maps of the axisymmetric radial and longitudinal components of the solar magnetic field have been obtained, extending over four cycles[97]. These two components can be obtained by many observatories. Determining the missing latitudinal component from polar observations over the course of a solar cycle would complete our understanding of the solar global magnetic field structure.

**Key advancements enabled by polar observations**:

1) Doppler velocity measurements from polar viewpoints will allow for accurate inference of internal solar flow fields in high-latitude regions. This will help identify the location of toroidal field generation and provide critical constraints for solar dynamo models.

2) High-precision magnetic field measurements in the polar regions will clarify the generation mechanism of the poloidal field, enabling the discrimination between different solar dynamo models.

### 2.2 What Drives the Fast Solar Wind?

#### 2.2.1 Key unresolved issues regarding the fast solar wind

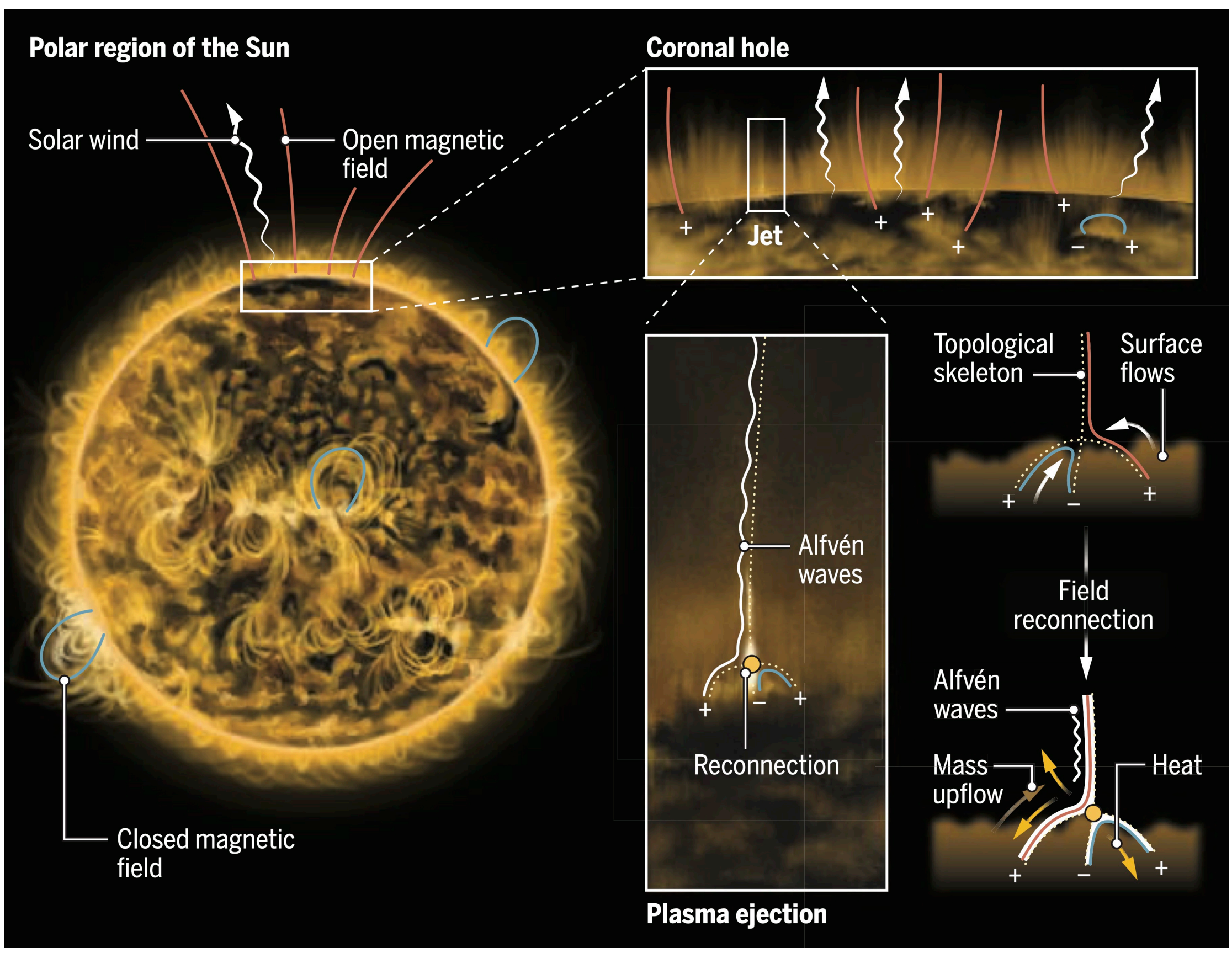

Figure 5 Solar polar observations by EUI on the Solar Orbiter spacecraft reveal plasma ejections (jets) that permeate coronal holes (areas of open magnetic fields). Jets arise from regions of magnetic reconnection. This reconfiguration of the magnetic fields also generates heat, Alfvén waves, and mass upflows that, together with the jets, provide energy and momentum to the solar wind. Figure from Ugarte-Urra et al[98].

The solar wind is a magnetized plasma stream originating from the solar atmosphere, gaining energy and becoming supersonic near the Sun. It flows outward from the solar corona, and extends throughout the solar system. The fast solar wind, often originating from polar coronal holes (see Figure 5)[98], permeates the majority of the heliospheric volume, dominating the physical environment of interplanetary space. Therefore, the fast solar wind is a key component of space weather and can significantly impact the Earth and its environment[99]. Its properties and origin remain central topics in heliospheric physics.

The Ulysses mission (1990–2009), the only spacecraft to have explored the heliosphere from a solar polar orbit, provided in-situ measurements of the solar wind plasma properties, the electromagnetic fields, and the ionic composition. Ulysses confirmed the bimodal nature of the solar wind around solar minimum[100]. This bimodal state consists of a relatively homogeneous fast solar wind (>500 km $s^{-1}$) at high heliolatitudes and a highly variable slow wind (<500 km $s^{-1}$) at lower latitudes. During the solar minimum, the fast solar wind originates from the polar coronal holes, which are dominated by single polarity and open magnetic fields[41], as shown in Figure 5. In contrast, the slow solar wind is largely confined to the space around the solar equatorial plane. The fast solar wind originates deep within large coronal holes[101], while the slow solar wind likely stems from open-closed magnetic field boundaries (e.g., active region edges, coronal hole boundaries), small low-latitude coronal holes, or helmet streamers[102–104].

In recent decades, significant progress has been made in understanding the origin and acceleration of the slow solar wind through in-situ measurements and high-resolution remote-sensing observations. However, the lack of direct imaging observations from high latitudes has hindered the studies of the fast solar wind. To date, the spectroscopic observations have identified the presence of coronal outflows within polar coronal holes[32,33,35,105]. However, the precise locations and altitudes within polar coronal holes where the fast solar wind originates remain unclear.

Coronal holes contain numerous ray-like polar plumes that extend over several solar radii[106]. Nevertheless the debate is ongoing regarding whether the fast solar wind originates from the plumes or the interplume regions. Some solar physicists believe that the polar plumes are the main source of material for the fast solar wind[107], while others argue that the contribution of the polar plumes is minimal and may even be negligible[108]. The observations obtained by the Interface Region Imaging Spectrograph (IRIS)[109] and Solar Orbiter have detected ubiquitous small-scale jets (with velocities on the order of 100 km $s^{-1}$) within coronal holes, which may supply sufficient hot plasma to sustain the solar wind[22,25]. A recent study of the solar wind from PSP revealed a range of wind speeds, with the lower end matching the classical Parker solar wind and the upper end consistent with the measurements of the jets/jetlets superimposed on the background Parker solar wind[21]. This suggests that the jets or jetlets could be key to understanding solar wind generation. However, their precise role in mass and energy transport remains uncertain. Additionally, since most observations are restricted to the ecliptic plane, investigations of these structures suffer from significant observational biases.

Recently, two of the remote-sensing instruments aboard Solar Orbiter, EUI and SPICE, have provided new insights into the dynamics of the solar atmosphere[25,56,110]. Despite this progress, the existing observations in the ecliptic plane have been insufficient to definitively determine the origin of the fast solar wind, which remains to be a key unanswered question in solar and space physics. Magnetic reconnection[33,111,112] and wave-driven processes[100] have been proposed as potential drivers. Both mechanisms are believed to contribute to heating and accelerating of the solar wind[30,113], yet their relative importance and interplay are still debated due to the absence of direct observational constraints from polar vantage points. Transverse waves have been detected in solar polar regions, and they are hypothesized to provide a significant energy source for the acceleration of the fast solar wind. By combining the in-situ measurements obtained by PSP and Solar Orbiter, Rivera et al.[59] computed that the damping and mechanical work performed by the Alfvén waves are sufficient to power the heating and acceleration of the fast solar wind in the inner heliosphere.

However since the measurement by PSP was obtained at 13.3 solar radii, it is still unknown what accelerates the fast solar wind from its source region.

Large polar coronal holes, the primary source regions of the fast solar wind, may host distinct wind generation processes compared to the mid- and low-latitude coronal holes. However, this hypothesis remains untested due to the observational limitations. In-situ measurements near the ecliptic plane typically detect fast wind streams originating from the polar coronal hole boundaries or the mid-/low- latitude coronal holes [113]. Magnetic connectivity analyses suggest that the steady fast solar wind from polar coronal holes rarely reaches the detectors in the ecliptic plane. The Ulysses observations revealed that the fast solar wind from the polar hole interiors exhibits remarkable stability, whereas streams from the polar hole boundaries or mid-latitude holes display significant temporal variability, suggesting potentially distinct origin and acceleration processes[41].

### 2.2.2 Limitations in current research

Backtracking methods are usually employed to locate the source region of the solar wind, which trace solar wind streams measured in-situ by spacecrafts back to their solar surface origins. These methods rely on magnetic field extrapolations or global magnetohydrodynamic (MHD) simulations[114]. However, past applications have primarily focused on observations near the ecliptic plane, where interactions between fast and slow solar wind streams introduce significant uncertainties[115,116]. Because the slow solar wind generally originates near the ecliptic, in-situ observations in the ecliptic plane often detect complex interactions between the fast and slow streams, complicating backtracking efforts. Specifically, the fast solar wind streams overtake and compress the preceding slow streams, forming CIRs. As the heliocentric distance increases, the mix between the fast and slow solar wind becomes even more pronounced[117], further obscuring the source region identification.

To unveil the origin of the fast solar wind, remote-sensing observations of photospheric magnetic fields and coronal emission structures in the polar regions are essential. However, current observations are limited in the ecliptic plane, where the projection effect and other limitations severely degrade measurement accuracy and reliability. In the absence of continuous, high-precision magnetic field observations of the polar regions, researchers have to extrapolate the low-latitude magnetic field to estimate the polar magnetic fields. This method is highly assumption-dependent and prone to systematic errors. Notably, the extrapolated open magnetic flux falls below the values obtained from in-situ heliospheric measurements[118]. This observational deficiency fundamentally restricts progress in understanding the fast solar wind origin. Additionally, remote-sensing of polar coronal structures from the ecliptic plane suffers from foreground contamination by mid- and low-latitude emission structures, obscuring crucial spatial and altitudinal information needed to pinpoint the origin of the fast solar wind.

Prevailing theories link fast solar wind acceleration to either MHD wave dissipation or magnetic reconnection[22,25,30,33,100,109,111,112]. However, due to foreground contamination it is not easy to accurately characterize MHD waves and magnetic reconnection events within polar coronal holes from oblique ecliptic viewpoints. An overhead view provides direct and unobstructed visibility of the entire polar region, offering a clearer advantage for identifying and characterizing such dynamic processes. Furthermore, the lack of precise polar photospheric magnetic field measurements limits the reliability of the numerical simulations of the polar dynamics. It also remains unclear whether the empirical relationships derived from the mid- and low-latitude coronal holes can be reliably extended to the large polar coronal holes, which are believed to host distinct wind acceleration processes.

### 2.2.3 The need for polar observations

Current research on the origin of the fast solar wind primarily relies on observations from near the ecliptic plane. Although the Ulysses mission has provided valuable in-situ measurements of the fast solar wind at high heliolatitudes, it lacked remote-sensing capabilities to directly image its source regions and ceased operations in 2009. Direct remote-sensing observations of the Sun's polar regions from high heliolatitudes, combined with in-situ measurements, can provide crucial insights into the mass and energy sources for the fast solar wind, and help distinguish between different mechanisms that drive the fast solar wind, such as MHD waves and magnetic reconnection. Such observations would resolve critical uncertainties of the dynamics in the coronal hole, refine solar wind models, and enhance our understanding of the Sun's open magnetic flux and its heliospheric impact.

**Key advancements enabled by polar observations**:

1) Observations of the fast solar wind from high heliolatitude viewpoints would circumvent the complications arising from the fast-slow stream interactions, significantly improving the accuracy of backtracking the fast streams to their source regions.

2) Direct imaging observations of the Sun's poles form high latitudes would eliminate the projection effect, providing comprehensive and accurate measurements of the spatiotemporal evolution of the magnetic fields, and emission structures within polar coronal holes. These measurements would provide crucial insights into the mass and energy supply to the fast solar wind. By quantifying the roles of the MHD waves and the magnetic reconnection, these observations would enable comparisons with the solar wind originated in the mid- and low-latitude coronal holes, revealing potential differences in their driving mechanisms.

## 2.3 How Do the Space Weather Processes Globally Originate and Propagate Throughout the Solar System?

### 2.3.1 Key unresolved issues regarding the global space weather processes

Heliospheric space weather refers to the disturbances in the heliospheric environment caused by the solar wind and solar eruptive activities. Extreme space weather events, such as large solar flares and CMEs (see Figure 6)[119,120], can significantly trigger space environmental disturbances such as severe geomagnetic storms, ionospheric storms, and changes in atmospheric density, as well as spectacular aurora phenomena, posing a serious threat to the safety of high-tech activities of human beings. Thus developing effective methods to predict the space weather events, and further mitigate or prevent these impacts is of great importance. However, the generation and propagation of the space weather events involve complex physical processes spanning multiple orders of magnitude in both temporal and spatial scales. The current solar observations obtained in the ecliptic plane are insufficient to comprehensively monitor these phenomena across such vast scales, limiting our ability to 1) map three-dimensional (3D) structures of solar wind disturbances, and understand their 3D propagation and evolution throughout the heliosphere, and 2) obtain the 3D distributions of the energetic particles and magnetic fields in the heliosphere.

The global structure of the solar wind disturbance is shaped by both the internal solar processes and the atmospheric dynamics. Its propagation and evolution involve complex physical mechanisms, many of which remain poorly understood. The developments of solar wind models have evolved from simple analytical frameworks to sophisticated numerical simulations, with 3D MHD models serving as essential tools. These numerical models rely on the photospheric magnetic field observation as boundary conditions to reproduce 3D structures and evolution of the solar wind disturbances. However, the absence of the magnetic field measurements in the polar regions leads to significant uncertainties in the synoptic magnetic maps, which introduce errors in the coronal models, ultimately reduce the accuracy of the solar wind disturbance forecasts.

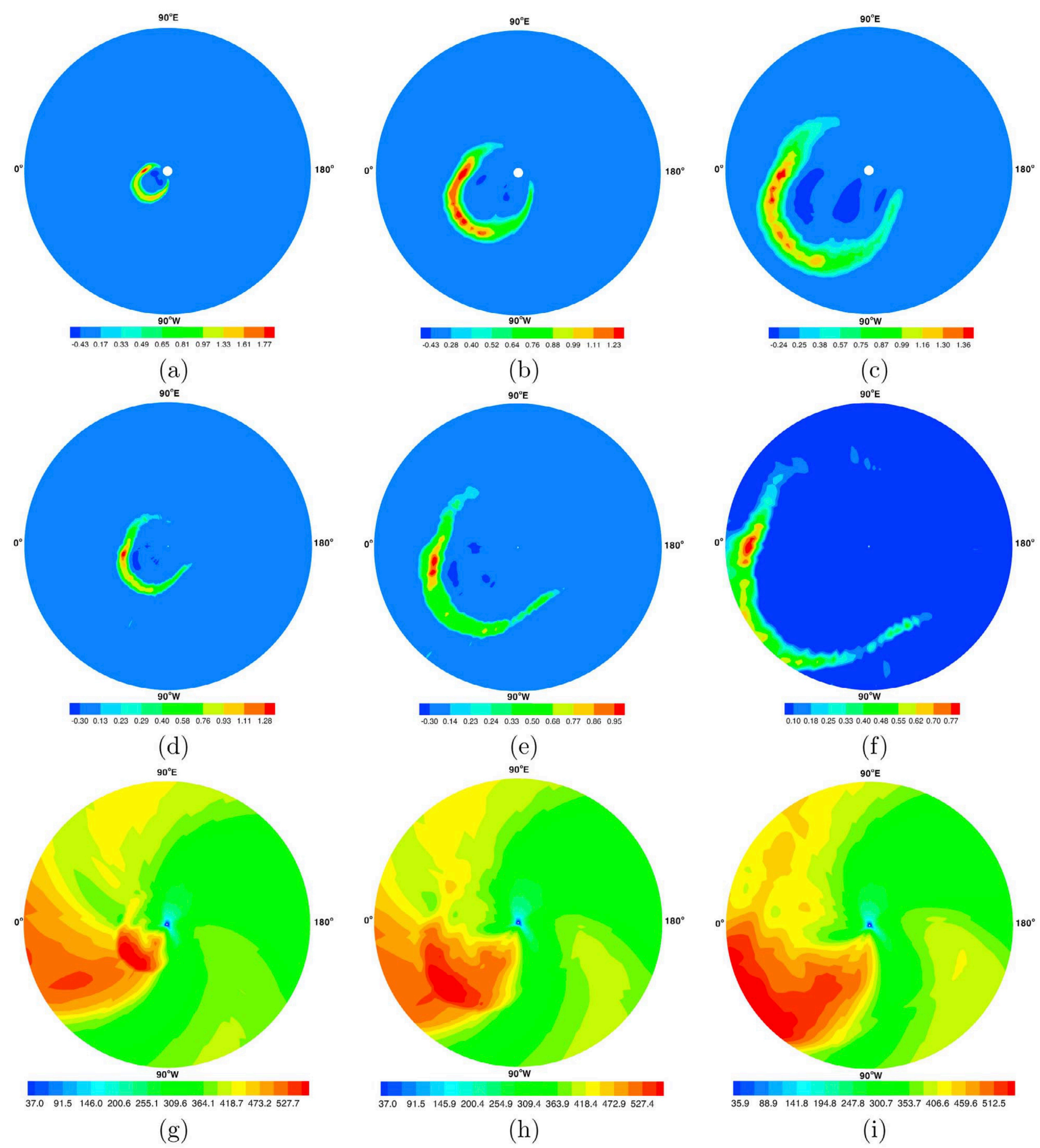

Figure 6 Simulated images of the relative density (in the unit of $cm^{-3}$, Panels (a) – (f)) and the velocity (in the unit of km $s^{-1}$, Panels (g) – (i)) during the propagation of a CME in the x-y azimuthal plane at (a) t = 1 hour, (b) t = 3 hours, (c) t = 5 hours, (d and g) t = 20 hours, (e and h) t = 40 hours, and (f and i) t = 60 hours. Panels (a)– (c) show a field of view from $1R_\odot$ to $30R_\odot$, and Panels (d)–(i) show a field of view up to 1 AU. Figure from Zhou et al[119].

Transient events like CMEs significantly alter the global distribution of the solar plasma and magnetic fields. Their high-density plasma and strong magnetic fields inject substantial mass and energy into interplanetary space, causing heliospheric disturbances. Current forecasting relies on remote-sensing observations of the activities and in-situ measurements of eruptive events near the ecliptic plane. However, these approaches lack global coverage of key processes, including solar atmospheric dynamics, CME initiation, and interplanetary propagation. Bridging these observational gaps is critical for improving space weather forecasting and understanding the global impact of solar eruptions on the heliosphere.

### 2.3.2 Limitations in current research

Heliospheric space weather forecasting focuses on predicting key phenomena, such as CIRs, CMEs, solar energetic particle (SEP) events, and their planetary impacts. Monitoring and modeling the solar wind throughout interplanetary space are essential for predicting these events and mitigating their adverse effects.

CIRs are particularly important in solar wind dynamics. Coordinated in-situ measurements from the Solar Terrestrial Relations Observatory (STEREO)[121] and ACE spacecraft have allowed for tracking CIR propagation and evolution within the ecliptic plane over multiple days. However, the lack of multi-angle observations, particularly from out-of-ecliptic vantage points, limits the accurate reconstruction of their 3D structure and dynamics. CIRs may exhibit entirely different morphologies when observe from different heliolatitude viewpoints. Direct imaging observations out of the ecliptic plane could easily identify the spiral morphology of CIRs and detect multiple coexisting CIRs. This enables the continuous tracing of an interplanetary CIR back to its coronal source[122].

CMEs also present challenges for heliospheric forecasting. Observations have shown that CMEs do not always follow a straight trajectory through the corona and inner heliosphere, frequently undergoing significant deflections[123]. Therefore, determining the propagation trajectory of a CME within the ecliptic plane is crucial for determining whether and when the CME will hit the Earth. Most solar wind monitors, such as those at the Sun-Earth L1 point, cannot resolve the 3D structure of CMEs. While STEREO provides multi-perspective imaging, its ecliptic orbit limits observations of longitudinal CME deflections, morphological changes, and velocity evolution. White-light imaging observations from high heliolatitude viewpoints would allow for continuous CME tracking from the coronal initiation, improving measurements of the propagation speed, longitudinal span, and morphological evolution[124]. In addition, multi-perspective observations are essential for 3D localization and reconstruction, significantly reducing the projection effects and model fitting uncertainties[125].

The accuracy of the CME simulations is also significantly influenced by the background solar wind conditions[126,127]. Incomplete knowledge of polar magnetic fields leads to gaps in global synoptic magnetic maps, weakening the boundary conditions required for MHD models. This degrades the simulations of both steady-state solar wind and transient CME evolution, limiting predictive capabilities.

SEP events, posing serious hazards to spacecraft, astronauts, and aviation, are another challenge in space weather forecasting. The onset and intensity of SEP events are strongly correlated with the CME dynamics[128]. However, insufficient stereoscopic monitoring of the shock source regions, combined with limited knowledge of polar magnetic field configurations, hinders accurate modeling of shock formation and particle acceleration processes. Furthermore, the absence knowledge about the magnetic fields in the polar regions limits the shock modeling, reducing our ability to accurately study shock-driven particle acceleration.

### 2.3.3 The need for polar observations

Currently, heliospheric observations are largely limited in the ecliptic plane, preventing a full 3D characterization of CMEs, shocks, and SEP events. This observation gap undermines the accuracy of space weather forecasting for the Earth and other planets. Observations obtained by ground-based instruments and ecliptic-plane spacecraft fail to resolve the dynamics in the polar regions, and the lack of polar magnetic field observations introduce significant uncertainties into global MHD models due to the ill-constrained boundary conditions.

These limitations restrict progress in both heliospheric modeling and the prediction of space weather impacts. To overcome them, high-latitude solar observations are essential. A mission designed to observe the Sun from polar vantage points, particularly combining with observation carried out from the ecliptic plane, would deliver unprecedented stereoscopic views of solar wind disturbances and their 3D propagation. Such synergistic observations would benefit the development of data-driven, time-dependent 3D MHD models with comprehensive boundary conditions. These models will also be validated through coordinated multi-wavelength remote-sensing and in-situ measurements. Ultimately they could unveil the global structures of the solar plasma and the magnetic fields, enabling a quantitative mapping of the solar wind disturbances and their 3D evolution and propagation throughout the heliosphere.

**Key advancements enabled by polar observations**:

1) Observations from high heliolatitude viewpoints, combined with the observations obtained in the ecliptic plane, enable better constrained global simulations of the 3D evolution and propagation of solar wind disturbances and CMEs throughout the heliosphere.

2) Multi-angle EUV and X-ray imaging observations of the CME initiation, combined with coronagraphic observations, would better constrain CME models and their propagation.

3) Remote-sensing and in-situ measurements of the solar wind and CMEs from polar vantage points would provide critical insights into high-latitude solar wind dynamics.

4) Coordinated monitoring of the solar eruptions across latitudes would advance understanding of particle acceleration and diffusion processes.

# 3 Proposed Missions to Explore the Sun's Poles

While the Solar Orbiter mission progresses and ventures to higher solar latitudes, the max heliolatitude that the Solar Orbiter can reach is limited to 34°. That said, it still falls short of providing enough details to understand the role of the polar regions in driving the solar magnetic cycle and the fast solar wind.

Recognizing the critical need for the observations of the Sun's poles from high heliolatitude viewpoints to address the scientific questions discussed in Section 2, several mission concepts have been proposed, including: the Solar Polar Imager (SPI)[129,130], the POLAR investigation of the Sun (POLARIS)[131], the Solar Polar ORbit Telescope (SPORT)[132,133], the Solaris mission[134], the High Inclination Solar Mission (HISM)[135], and SPO etc.

SPI was proposed to utilize a solar sail propulsion to reach a 0.48 AU circular orbit with an inclination of 75°, enabling high latitude observations of the solar poles[130]. The mission planned to carry five remote-sensing instruments and three in-situ instruments aimed at exploring the connections between the Sun, solar wind, and solar energetic particles.

The POLARIS mission was proposed to use a combination of a gravity assist and solar sail propulsion to place a spacecraft in a 0.48 AU circular orbit around the Sun with an inclination of 75°[131]. The mission aimed at advancing the understanding of solar magnetic cycle, 3D structure below the surface, solar wind, and the acceleration and transportation of the solar energetic particles, using five remote-sensing instruments and four in-situ instruments.

The SPORT mission was initially proposed in 2004[132] and later was selected for intensive scientific and engineering background studies under the Chinese Space Science Strategic Pioneer Project since 2011[133]. It was designed to reach an orbital inclination of no less than 60° through multiple gravity assists from Jupiter and Earth (or Venus). The mission planned to carry multiple remote-sensing and in-situ instruments, with scientific objectives 1) to reveal the physical processes of CME propagation and evolution in the inner heliosphere, 2) to uncover the relationships between high-latitude solar magnetic activity, solar eruptive phenomena, and the solar activity cycle, 3) to understand the acceleration, transport, and distribution of energetic particles in the corona and heliosphere, and 4) to unveil the origin and physical characteristics of the fast solar wind. Although SPORT did not proceed to satellite engineering phase after completing the scientific and engineering background studies, it provided valuable scientific and engineering groundwork for several subsequent satellite mission proposals.

The Solaris mission was proposed to use a Jupiter Gravity Assist (JGA) to raise the orbital inclination to 75°, followed by multiple Venus Gravity Assists (VGAs) to gradually reduce its orbit's aphelion. The mission planned to carry three remote-sensing instruments: the Compact Doppler Magnetograph (CDM), the EUV Imager (S-EUVI) and the White Light Coronagraph (S-COR), and three in-situ instruments: the Magnetometer (MAG), the Ion-Electron Spectrometer (IES), and the Fast Imaging Plasma Spectrometer (FIPS)[134]. Although Solaris received a NASA Phase A support for concept study in 2020, it was not selected for flight in 2022.

The HISM mission concept is largely based on SPI, employing the solar sail techniques. It is designed to take approximately 2.6 years to venture to a 0.48 AU ecliptic orbit, after which it will increase the orbital inclination at a rate of around 10° per year[135]. Ultimately HISM will reach an orbital inclination larger than 75°. The mission plans to carry two remote-sensing instruments and five in-situ instruments aimed at investigating the solar wind and space weather processes.

These proposed missions highlight the growing recognition of the importance of solar polar observations in advancing heliophysics. By bridging the observational gap left by ecliptic-based missions, these missions can provide critical insights into the solar magnetic cycle, solar wind origin, and space weather prediction.

# 4 The SPO Mission

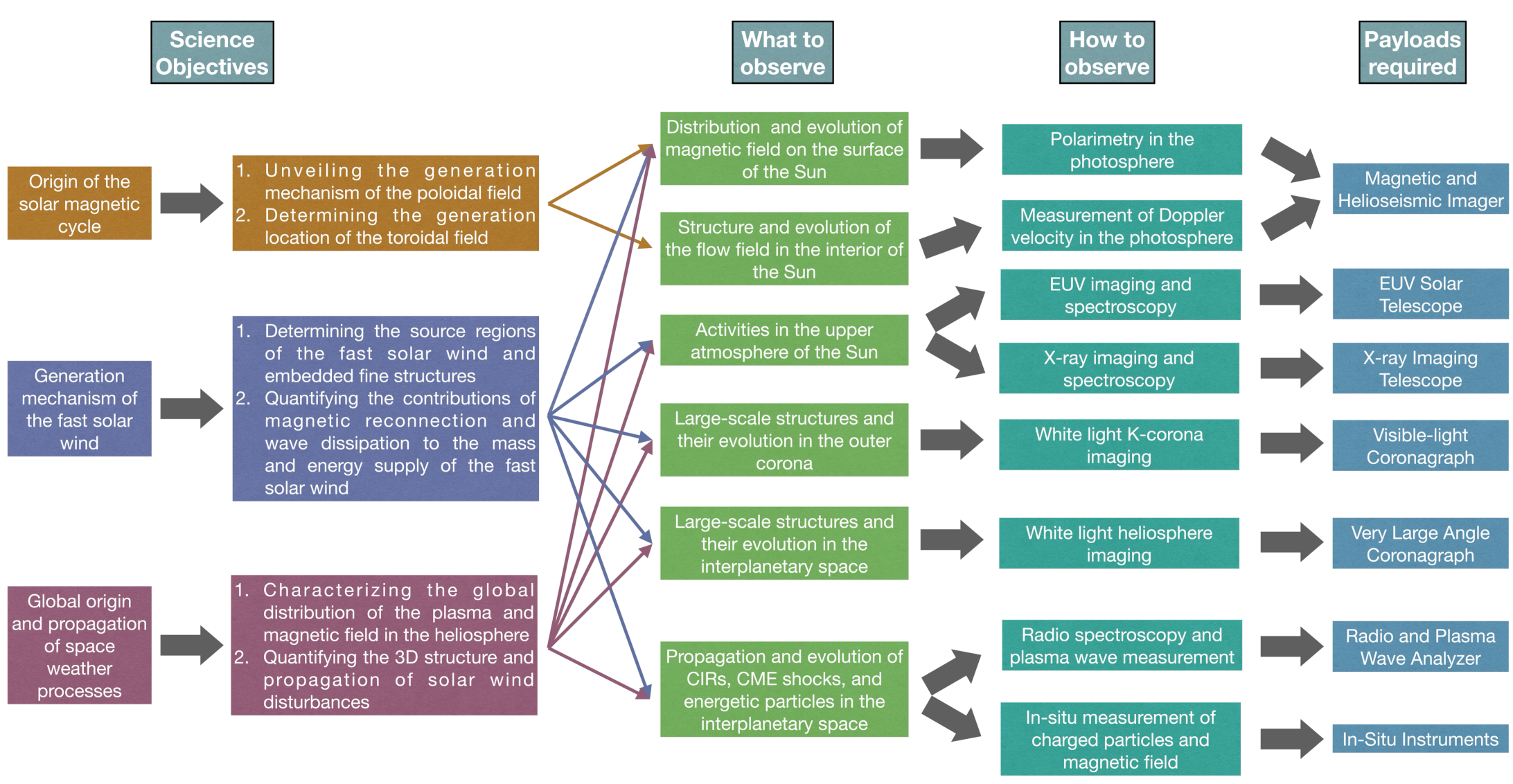

Figure 7 The scientific goals of the SPO mission and the corresponding payload requirements.

The SPO mission, aiming at the breakthrough on the top-level scientific questions discussed in Section 2, is scheduled to be launched in January, 2029. Using a JGA, SPO will reach a maximum heliolatitude of ~ 75° (with up to 80° in an extended mission) to provide images of the Sun's poles. After several Earth gravitational assists, SPO will settle into an orbit with a period of approximate 1.5 years and a perihelion of approximately 1 AU. Designed for a 15-year operational lifetime (including the extended mission phase), SPO will observe both solar maximum and minimum, covering a complete solar cycle. During the first eight years of the mission, SPO will fly over the north and south poles once each. In the subsequent seven-year extended mission, it will fly over each pole four times. The orbit allows a ~300-day observation at latitudes larger than $55^{\circ}$ during the first north (June-November 2034) and south (March-July 2035) polar passes. Considering the fact that the next solar maximum will likely happen around 2035, the chance of SPO to capture the polar field reversal, a hallmark event in the solar cycle, is high.

SPO will complement ongoing solar missions, i.e. the STEREO Mission, the Hinode satellite, the Solar Dynamics Observatory (SDO)[136]. IRIS, the Advanced Space-based Solar Observatory (ASO- S)[137], the Solar Orbiter, the Aditya-L1 mission[138], the PUNCH mission[139], as well as the upcoming L5 missions (e.g., ESA's Vigil mission[140]). Although many of these missions may phase out, new missions will be put into operation when SPO passes over the Sun' Poles. By integrating observations from these missions, SPO will significantly contribute to an unprecedented "4π-observation" of the Sun.

In order to achieve its scientific goals, SPO will carry six remote-sensing instruments and four in-situ instruments to measure the vector magnetic fields and Doppler velocity fields, to observe the Sun in EUV, X-ray, and radio wave bands, to image the corona and the heliosphere up to 45 $R_{\odot}$, and to perform in-situ measurements of the magnetic field, and low- and high-energy particles in the solar wind.

A summary of the scientific goals of the SPO mission and the corresponding payload requirements is provided in Figure 7. The following subsections outline the capabilities of each instrument and how they are designed to address the scientific questions discussed in Section 3.

### 4.1 The Magnetic and Helioseismic Imager (MHI)

The Magnetic and Helioseismic Imager (MHI) is designed to measure the magnetic fields and Doppler velocities in the solar photosphere. By providing imaging observations from a polar vantage point, MHI will fill the observational gap in the magnetic and velocity field observations of the polar regions, which is essential for all three scientific objectives of the SPO mission.

Table 1 Observational requirements for MHI.

| Parameters | | Specifications |
|---|---|---|
| Telescope | Spectral line | Fe I 5324 Å |
| | FOV | 34′ |
| | Spatial Res. | 2″ |
| Longitudinal Magnetic field | Cadence | 60 s |
| | Sensitivity | 20 G |
| Vector Magnetic field | Cadence | 12 min |
| | Sensitivity | 200 G |

| Doppler Velocity | Cadence | 60 s |
|---|---|---|
| | Sensitivity | 100 m s$^{-1}$ |

To answer the scientific question, understanding how the solar dynamo works and drives the solar magnetic cycle, two key aspects must be addressed: 1) to pinpoint the location where the toroidal magnetic field is formed, and 2) to determine the primary mechanism behind the generation of the poloidal field. The measurements of the magnetic and velocity fields are crucial to address these questions. The measurements of the magnetic fields on the entire solar disk including the polar regions, active regions, and network regions are essential for understanding the generation mechanism of the poloidal field. Additionally, measuring velocity fields in the polar regions enables helioseismic inversions to map solar interior flow fields, which is key to determine the location of the toroidal magnetic field generation.

To ensure high-accuracy flow field measurements, the noise from (super)granulation must be minimized. Since polar regions may contain fewer supergranules than low-latitude regions, long-duration observations are required to average out supergranulation signals[141]. For precise magnetic field measurements, MHI must resolve the network magnetic field structure, requiring a spatial resolution better than 2″ to distinguish network field structures, a longitudinal magnetic field sensitivity below 20 G to detect weak network magnetic fields, and a transverse magnetic field sensitivity below 200 G to detect strong horizontal magnetic field structures within the network regions.

In addition to its contribution to the solar dynamo science, MHI also plays a key role in achieving the second scientific objective of the SPO mission, i.e. determining the origin of the fast solar wind. The debate over whether the fast solar wind originates from the polar plumes or the interplume regions[106,107] remains unresolved. Common methods for localizing the source region of the solar wind involves the global magnetic field extrapolation[142] and the global data-driven MHD simulations, which suffers from uncertainties due to the absence of accurate measurements of the polar magnetic fields. MHI will provide high-accuracy vector magnetic field measurements in the polar regions, overcoming these limitations. By combining with ecliptic plane observations, it will enable the construction of high-resolution global synoptic magnetic field maps, which serve as the boundary conditions for solar wind tracing, and the key inputs for numerical simulations of the solar wind and CME propagation throughout the heliosphere, which is also essential for the third scientific objective of the SPO mission, i.e. characterizing the global origin as well as the heliospheric propagation of space weather processes. The observation requirements for MHI to achieve these scientific goals are summarized in Table 1.

### 4.2 The Extreme Ultraviolet Telescope (EUT)

The Extreme Ultraviolet Telescope (EUT), consisting of the Multi-band EUV Imager (MEI) and the Full-disk Integrated Spectrograph (FIS), will provide crucial information to address the second and third scientific objectives of the SPO mission. Ideally, an EUV slit spectrograph capable of measuring LOS and turbulent velocities in the transition region and corona would be included. However, due to resource limitations on this deep-space mission and the technological difficulty in the mission team, a slit spectrograph is currently not onboard. Nevertheless, MEI and FIS still will significantly enhance our understanding of the fast solar wind and global space weather processes in the heliosphere.

MEI provides high-resolution images of the fine structures and their rapid evolution within coronal holes, including coronal bright points, jets, spicules, plumes, and interplume structures, which are the potential source regions of the fast solar wind. Observing these structures from a high heliolatitude is essential for identifying the origin site (location, height) of the fast solar wind, and determining the role of magnetic reconnection and MHD wave dissipation in supplying mass and energy to the fast solar wind.

Table 2 Observational requirements for EUT.

| Parameters | MEI | FIS |
|---|---|---|
| FOV | $\geqslant 48'$ | $\geqslant 40'$ |
| Spatial Res. | $\leqslant 2''$ | - |
| Detector size | ⩾ 4k×4k | 2k×1 |
| Wavelength | 17.1, 19.3, 30.4 nm (±0.1 nm) | 18.3–21.3 nm |
| Band width /Spectral Res. | ⩽0.7 nm @17.1 nm,⩽0.8 nm @19.3 nm ⩽1.5 nm @30.4 nm | ⩽0.04 nm @20 nm |
| Cadence | ⩽1 min | ⩽5 min |

MEI also contributes to the third scientific objective by monitoring the origin, evolution, and early propagation of the solar eruptions, providing critical constraints for the 3D evolution and propagation of CMEs in the heliosphere. This is essential for establishing global numerical simulations of the solar wind disturbances and improving the prediction of space weather events.

FIS is designed to characterize the latitudinal distribution of the solar EUV radiation and the coronal physical properties. It is also able to detect the LOS velocities of CMEs during their early propagation[143,144], providing essential constraints for the global space weather modeling, and improving the predictions of CME propagation and its impact on Earth and other planets. Additionally, FIS is able to provide unique reference for explorations of stellar atmospheres and their eruptions, as solar observations at different phases of SPO's orbit are basically similar to observations of stars with different rotational-axis inclinations.

The field of view (FOV) of MEI should be at least $48'$, enabling observations of the entire solar disk and atmospheric structures within 0.5 $R_\odot$ beyond the solar limb, covering typical energy release heights of solar eruptions. The selected wavelength bands include 17.1 nm (~$9\times10^5$ K), 19.3 nm (~$1.6\times10^6$ K), and 30.4 nm (~$5\times10^4$ K) to cover a broad temperature range (from the chromosphere to the corona). The spatial resolution of MEI should be better than 2″, and the temporal resolution should be one minute, meeting the observational requirements for EUV emission structures such as network structures, polar plumes, coronal bright points, jets, and filaments.

The 18.3–21.3 nm band is adopted by FIS to cover a temperature range from $10^{5.6}$ K to $10^{6.3}$ K, the typical temperature range of coronal and CME plasma. Moreover, the Fe VIII-XIV lines in these wavelength bands can be used for temperature diagnostics, and the lines of the Fe XII at 19.512 and 18.688 nm can be used for density diagnostics. Several strong Fe lines could also be used for the velocity measurements of CMEs during their early propagation stage. The spectral resolution might need to be better than ~500[143-145], i.e. better than 0.04 nm at 20 nm, to achieve enough accuracy of the LOS velocity measurement. The FOV of at least 40′ is required to cover major EUV structures extending beyond the solar limb (typically within 1.25 $R_\odot$). A detailed summary of the observational requirements for MEI and FIS is provided in Table 2.

### 4.3 VISible-light CORonagraph (VISCOR)

The VISible Light CORonagraph (VISCOR) is designed to image the white-light corona up to at least 6.0 $R_\odot$, capturing its intensity and polarization. This capability directly contributes to the second and third scientific objectives of the SPO mission.

VISCOR will provide high-resolution imaging observations of the coronal structures such as the plumes, interplume regions, and streamers from viewing vantage points out of the ecliptic plane, offering crucial insights into the origin and acceleration of the fast solar wind[146–149]. Additionally, VISCOR will significantly improve the ability to constrain the 3D global models of the solar wind, enabling tracing the solar wind streams back to their source regions and quantifying wind super-radial expansion.

Moreover, the observations from high heliolatitudes provide a unique view of the propagation and evolution of CMEs in the ecliptic plane, particularly the Earth-directed CMEs, which may be difficult to detect from traditional viewpoints. VISCOR will thus help to improve the prediction of the space weather in the solar system, i.e. the third scientific objective of SPO. Furthermore, by combining with the observations obtained in the ecliptic plane, it is capable of constraining the 3D structures of CMEs and their associated shocks, allowing for better determination of the CME front, core, and shock positions, ultimately enhancing our understanding of the CME evolution and improving the ability to model and predict CME paths and their impacts on the solar system environment.

Table 3 Observational requirements for VISCOR.

| Parameters | Specifications |
|---|---|
| Waveband | 540–640 nm |

| FOV | 0.4° – ⩾1.6° (effective 2.2–⩾6.0 $R_\odot$ @1 AU) |
|---|---|
| Spectial Res. | ⩽ 16″ |
| Linear polarizer positions | 0°, ±60° |
| Time cadence | ⩽6 min(CME mode ⩽2 min)<br>Specific ~30 s Obs. for short period |
| Stray-light level | ⩽ $5 \times 10^{-11}$ MSB @1.6° |

VISCOR is a classical externally occulted coronagraph, similar to SOHO/LASCO-C2[150], but with a larger aperture, higher quantum efficiency, and higher cadence. The FOV ranges from 0.4° to at least 1.6° ( effective 2.2 $R_\odot$ to 6.0 $R_\odot$ at 1 AU) with respect to the disk center, which is necessary for imaging the corona, and tracking the acceleration and propagation of the solar wind and the CMEs. The goal of the VISCOR stray-light level is expected to be better than $5 \times 10^{-11}$ MSB at 1.6°, enabling the detection of weak coronal features like faint plumes or shocks far from the solar disk. The spatial resolution should be better than 16″ to resolve the plume and inter-plume regions, the streamers, the switchbacks, and the CME-related structures. The regular time cadence for the solar wind observations should be less than 6 min, with the capacities of high cadence modes of 30 sec for the coronal density fluctuations, and 2 min for the CME observations, which are vital for detecting coronal dynamics and tracking fast-evolving CMEs. A detailed summary of the observational requirements for VISCOR is provided in Table 3.

### 4.4 The Very Large Angle CORonagraph (VLACOR)

As a complement to VISCOR, the Very Large Angle CORonagraph (VLACOR) is designed to observe the heliosphere with a larger FOV to effectively image the large-scale structures and transients in the solar wind and to track the propagation of CMEs and their corresponding shocks, which are crucial for the second and third scientific objectives of the SPO mission.

Table 4 Observational requirements for VLACOR.

| Parameters | Specifications |
|---|---|
| Aperture | 12 mm |
| FOV (Half angle) | 1.33° – 12° |
| | 5–45$R_\odot$ @1AU |
| Pixel Res. | 45″/pixel |
| Image array | 2k×2k,14bit A/D |
| Spectral bandpass | 600–750 nm |
| Cadence | Brightness: 15 min |
| Stray light | ⩽$7\times10^{-10}$ MSB@5$R_\odot$ |
| | ⩽$1\times10^{-12}$ MSB@45$R_\odot$ |

Large-scale heliospheric structures, such as solar wind, coronal streamers and CMEs, can extend across several solar radii. As the solar wind evolves, it undergoes acceleration from subsonic to supersonic speeds. In the inner heliosphere beyond 10 $R_\odot$, the solar wind continues to accelerate[151], but the acceleration mechanisms remain debatable. Moreover, forward modeling of the Thomson scattering of 3D MHD solar wind simulations reveals that coronal and interplanetary structures and transients, such as CIRs and CMEs, exhibit distinct morphologies when observed from high-latitude vantage points[124]. White-light imaging observations outside the ecliptic plane at a latitude exceeding 60° can clearly identify the spiral morphology of CIRs and detect multiple coexisting CIRs. Combining with the imaging observation provided by MEI， it enables the continuous tracing of interplanetary CIRs back to their source regions[122]. Observing the evolution of these structures from the lower corona to the heliosphere will provide key insights into the origin of the fast solar wind, directly addressing the second objective of the SPO mission.

In addition, the white-light imaging observations of the heliosphere outside the ecliptic plane are capable of monitoring the CME propagation in the ecliptic plane from its coronal initiation, and then determining its propagating speed and longitudinal span[124]. Previous studies have shown that CMEs may not propagate along a straight trajectory through the corona and inner heliosphere, but undergo deflected propagation[123]. Identifying the precise propagation trajectory of a CME within the ecliptic plane is essential for predicting it potential Earth impacts. Additionally, directly imaging solar wind structures and transients from a polar vantage point provides critical constraints for data-driven global heliospheric models, improving the accuracy of space weather prediction.

The inner edge of the FOV of VLACOR is proposed to be 5 $R_\odot$ @1 AU, with an overlap of the FOV of VISCOR, which reaches 6 $R_\odot$ when the SPO spacecraft is 1 AU from the Sun. The outer FOV of VLACOR is designed to be as larger as possible within the limitation that a single instrument can reach, i.e. accordingly, 45 $R_\odot$ @1 AU. By combining VISCOR and VLACOR, SPO will provide comprehensive white-light observations, covering the primary regions where the fast solar wind is accelerated. Additionally, these observations will encompass the full acceleration phase of the CMEs, ensuring continuous tracking of their evolution. A detailed summary of the observational requirements for VLACOR is provided in Table 4.

### 4.5 The X-ray Imaging Telescope (XIT)

The X-ray Imaging Telescope (XIT) is designed to observe X-ray bursts and small-scale activities in the polar regions and coronal holes. Its scientific objectives are 1) to investigate the small-scale activities in the source regions of the fast solar wind, and 2) to study the energy release and the Hard X-Ray (HXR) directivity in X-ray bursts, which are closely related to the second and third objectives of the SPO mission, respectively. XIT consists of two instruments, an X-ray imager and a soft X-ray spectrometer. Together, these instruments can provide full-disk light curves, spectra, and images of X-ray bursts in the energy range of 5–100 keV, as well as the soft X-ray spectra in 0.5–10 keV, within a fixed, small FOV limited to about 5 arcmin. The main specifications of XIT are listed in Table 5.

Table 5 Observational requirements for XIT.

| Parameters | HXR | SXR |
|---|---|---|
| Telescope | 5–100 keV<br>spectra and images | 0.5–10 keV<br>spectra |
| FOV | >43′ | 4′ – 6′ |
| Spatial Res. | 6.5″ | - |
| Energy Res. | 5 keV @59.5 keV | 0.2 keV @6 keV |
| Time Res. | 4 s, burst 1 s (0.25–2 s) | 8 s (4–24 s) |
| Pointing Accuracy | Better than 2″ | - |
| Data Size | ~0.84 GB/d | ~0.021 GB/d |
| Size | 1000 mm×300 mm×300 mm | |
| Mass | 9.5 kg | |

The X-ray bursts, direct products of energy release in solar flares, are widely used to study plasma heating and energetic electrons[152–155]. The low-energy X-rays from the Sun are mostly emitted by the thermal plasma at temperatures ranging from a few to tens of MK. In contrast, at energies above ~20 keV, the X-ray emissions primarily result from the accelerated electrons interacting with ambient plasma. By analyzing the X-ray spectra and images, key parameters of the thermal and non-thermal electrons can be obtained, which are essential for understanding solar atmospheric dynamics during flares and constraining the global models of the solar eruptive events. More importantly, by combining the XIT observations with other X-ray instruments positioned at different viewing angles, we can explore the 3D properties of the X-ray sources, the directivity of the solar X-rays[156], and the pitch-angle distribution of the energetic electrons. To achieve these goals, the FOV should be larger than 42′ to cover the low corona, where the X-ray coronal sources are often found. The energy range should cover 5–100 keV to include thermal and non-thermal emissions. The time resolution should be better than 2 s to observe the typical variations of the flare HXR. To resolve the typical flare loops and double footpoints, the spatial resolution should be better than 7″. The spectral resolution should be better than 5 keV @59.5 keV for the HXR spectral analysis.

On the other hand, XIT can also help identify possible sources of the fast solar wind by observing soft X-ray emissions from the small-scale activities in the quiet regions and coronal holes, including X-ray bright points and small-scale jets, of which the X-ray signals have been observed and studied [157,158]. Although the imaging and imaging spectroscopy of these activities in Soft X-Ray (SXR) would be highly beneficial, the very limited resources on the deep space platform of SPO prevent the inclusion a full imager. Instead, XIT focuses on obtaining SXR spectra, which provide key plasma diagnostics, including temperature, emission measure, and elemental abundances. Since the platform always points to the solar disk center, the SXR detectors always observe a small region close to the disk center, waiting for any structures that pass through. We realize that observing coronal holes will not be a problem when the platform is over the polar regions during solar minimum, as the polar regions will be dominated by large coronal holes at solar minimum. Due to the relatively weak SXR emissions from these small-scale activities, an entrance window is implemented to restrict the FOV to 5 arcminutes, minimizing contamination from other solar regions. The energy range of SXR spectra should cover 0.5–10 keV, encompassing the thermal continuum and most of the emission lines. The energy resolution should be better than 0.3 keV @6 keV for quantitatively analysis of the spectra. Since the typical lifetimes of coronal bright points are in the range of minutes to tens of hours[159–162], the time cadence of SXR spectra should be better than 30 s.

### 4.6 The Radio Wave Analyzer (RWA)

The Radio Wave Analyzer (RWA) is designed to detect the solar radio emissions at low frequency range up to 30 MHz and to measure the local plasma waves. RWA aims at detecting the spectrum, time-variation and direction of solar radio bursts. It will measure the time-variation, amplitude and phase of radio waves from multiple directions in space. The main specifications of RWA are listed in Table 6.

RWA plays a crucial role in addressing key objectives of the SPO mission, particularly in unveiling the origin of the fast solar wind. The detection of Type III bursts and solitary wave emissions from high solar latitudes and polar regions will provide essential evidence for this investigation. In addition, the highly turbulent corona and the solar wind are prone to generate solitary wave radiation. PSP observations suggest that the solitary wave originates in the solar wind acceleration region. Their dynamic spectrum, detected by RWA, will provides important insights into the solar wind acceleration.

Table 6 Observational requirements for RPWA.

| Parameters | Specifications |
|---|---|
| Frequency | 10 kHz–30 MHz |
| Antenna | ⩾7.5 m, tripole |
| Spectral Res. | < 5%$f_c$ |
| Temporal Res. | ⩽ 1 s |
| Dynamic range | ⩾80 dB |
| Sensitivity | ⩽ 10nV/$\sqrt{\text{Hz}}$: 10 kHz–30 MHz |

Additionally, observations of Type II and Type III radio bursts will yield valuable clues on the energy release mechanisms in eruptive events, particle acceleration processes, the dynamic parameters of the solar wind, non-thermal high-energy particle flux characteristics, and the evolution of CMEs and interplanetary shocks. Based on the solar Type II and Type III observations of the polar regions, RWA will also provide diagnostic results of heliospheric plasma structures and densities above the polar regions.

### 4.7 The In-Situ Instrument Package

The in-situ instrument package consists of four instruments, including the Solar Wind Ion Mass Spectrometer (SWIMS), the Solar Wind Ion Retarding Potential Analyzer (SWIRPA), the Solar Energetic Particle Analyzer (SEPA), and the Magnetometer (MAG). This suite of instruments is primarily designed to measure and analyze key physical parameters of the solar wind, heliospheric energetic particles, and interplanetary magnetic fields. These measurements are crucial for charactering the states and evolution of the solar wind, energetic particles, and magnetic fields across various latitudes, as well as for understanding the impacts of solar activities on space weather throughout the solar system.

Table 7 Observational requirements for the in-situ Package.

| Instruments | Parameters | Specifications |
|---|---|---|
| SWIMS | Composition | Ions: H, $^{4}$He, $^{3}$He, C, N, O, Mg, Ne, Si, Fe |
| | Energy range | 0.2–80 keV/e |
| | Energy Res. | <10% |
| | Mass Res. | $(M/q)/\Delta(M/q) \geqslant 30$ @10 keV/e |
| | FOV | 360° × (±45° @10 keV/e) boresight of ±45° points to the Sun |
| | Angle Res. | 7.5° × 6° |
| | Geometric factor | 1 × 10$^{-5}$ cm$^{2}$ sr eV/eV |
| | Time Res. | 4 s (high time resolution mode) |
| | | 30 s (medium time resolution mode) |
| | | 300 s (low time resolution mode) |
| SWIRPA | Composition | Ions: H, $^{4}$He |
| | Energy range | 0.1–8 keV/e |
| | Energy Res. | <10% |
| | FOV | 90° cone toward Sun |
| | Angle Res. | 10° |
| | Geometric factor | 10 cm$^{2}$ sr |
| | Time Res. | Same as SWIMS |
| SEPA | Composition | Electrons and ions (H, $^{4}$He, C/N/O group, elements heavier than O) |
| | Energy Ranges | Electrons: 30 keV–5 MeV |
| | | Protons: 60 keV–100 MeV |
| | | Heavy ions: 12–100 MeV/n (Related to ion types) |
| | Energy Res. | <20% @ 200 keV (electrons & protons) |
| | Mass Res. | Ions (H, $^{4}$He, C/N/O, >O) |
| | FOV | Medium-Energy electrons<br>& High-Energy particles:2 units |
| | | Medium-Energy Protons:2 units |
| | Directional Res. | Medium-Energy electrons<br>& High-Energy particles: 50°/per direction |
| | | Medium-Energy Protons: 50°/per direction |
| | Geometric factor | >0.5 $cm^2$ sr |
| | Time Res. | Same as SWIMS |
| MAG | Measurement | ±128 nT, ±64000 nT |
| | Frequency range | DC~10 Hz |
| | Sample rate | 40 Hz |
| | Accuracy | better than 0.2 nT (±128 nT range) |
| | Noise | <0.01 nT/$\sqrt{\text{Hz}}$ @1 Hz |

By combining the interplanetary solar wind measurements with the photospheric magnetic fields obtained by the remote-sensing instrument MHI, the origin of the fast solar wind can be traced using techniques such as the Parker spiral model and global MHD simulations. This enables the identification of source regions in polar coronal holes, helping to address long-standing questions regarding the origin of the fast solar wind. Additionally, the in-situ measurement of interplanetary magnetic flux is essential for addressing the open magnetic flux problem. While the structure of the fast solar wind is less complex than that of the slow solar wind, it is not entirely uniform and may exhibit structural features originating from the Sun. Detailed in-situ measurements of these features by SWIMS, SWIRPA, SEPA, and MAG will help to identify the physical processes responsible for the generation of the fast solar wind.

Beyond providing insights into the origin of the fast solar wind, the in-situ measurements are critical for understanding the plasma composition and magnetic field topology of interplanetary coronal mass ejections (ICMEs), which are vital for studying the propagation and evolution of CMEs/ICMEs in the heliosphere and their potential impacts on the Earth. Additionally, shock waves driven by CMEs/ICMEs play a crucial role in accelerating particles to high energies, which can damage satellite hardware and disrupt radio communications. In-situ detection of these energetic particles is therefore essential to understanding how shocks accelerate particles and how these particles propagate through the heliosphere.

Given that the typical interplanetary magnetic field strength is around 6 nT at 1 AU and 0.3 nT at 5 AU, the MAG instrument must have an accuracy better than 0.2 nT to measure the weak interplanetary magnetic field around the SPO orbit. Furthermore, a time resolution on the order of sub-seconds is necessary to detect rapid magnetic field perturbations. To accurately characterize the fine features of the heliospheric energetic particle spectrum and its variations in response to changes in solar and interplanetary activity, the energy range should span from tens of keV to tens or even hundreds of MeV.

The energy resolution (dE/E) of SWIMS, SWIRPA, and SEPA should be better than 20%, with a time resolution on the order of minutes. The detailed observational requirements are summarized in Table 7.

# 5 Summary

Exploring the Sun and understanding how its activities impact throughout the heliosphere is one of humanity's greatest scientific endeavors. Over the past decades, solar observations have covered nearly the entire electromagnetic spectrum, encompassing X-rays, ultraviolet, optical, infrared, and radio waves, as well as the solar neutrinos, greatly enhancing our knowledge of the solar atmosphere and the dynamic processed therein. However, a critical observational gap remains, i.e. the inadequate observational coverage of the Sun's poles. Most solar missions have operated at the ecliptic plane, significantly limiting the ability to observe the solar poles. This observation gap raises three key scientific questions: 1) How does the solar dynamo work and drive the solar magnetic cycle? 2) What drives the fast solar wind? 3) How do space weather processes globally originate and propagate throughout the solar system? The most effective way to address these questions is through direct imaging of the Sun's poles and in-situ measurements of plasma and magnetic fields from high heliolatitudes.

The SPO mission has been proposed to address these three unanswered scientific questions. Its primary objectives are 1) to unveil the origin of the solar magnetic cycle, 2) to unveil the origin of the fast solar wind, and 3) to characterize the global origin and propagation of space weather processes. Equipped with six remote-sensing instruments and four in-situ instruments, SPO will, for the first time, capture images of the Sun's poles from high heliolatitudes. These groundbreaking observations will provide invaluable insights, revolutionizing our understanding of the Sun and the space weather processes.

## Acknowledgments

The forum was sponsored by ISSI-BJ. The financial and logistics support provided by ISSI, ISSI- BJ, and the National Space Science Center, Chinese Academy of Sciences are gratefully acknowledged. L.P.C. gratefully acknowledges funding by the European Union (ERC, ORIGIN, 101039844).